\documentclass[aps,twocolumn,groupedaddress,english,eqsecnum,showpacs,longbibliography]{revtex4-1}
\usepackage{amssymb}
\usepackage{amsmath}
\usepackage{graphicx}
\usepackage{xcolor}

\definecolor{green}{HTML}{229C70}
\definecolor{blue}{HTML}{23637B}
\usepackage[colorlinks,
            citecolor=blue,
            linkcolor=green,
            urlcolor=blue]{hyperref}

\usepackage[all]{hypcap} 

\makeatletter
\adddialect\l@English\l@english
\makeatother

\begin{document}
\title{Temperature dependence of the NMR relaxation rate $\mathbf{1/T_1}$ for quantum spin chains}
\author{Maxime Dupont}
\email{maxime.dupont@irsamc.ups-tlse.fr}
\affiliation{Laboratoire de Physique Th\'eorique, Universit\'e de Toulouse and CNRS, UPS (IRSAMC), F-31062, Toulouse, France}
\author{Sylvain Capponi}
\email{sylvain.capponi@irsamc.ups-tlse.fr}
\affiliation{Laboratoire de Physique Th\'eorique, Universit\'e de Toulouse and CNRS, UPS (IRSAMC), F-31062, Toulouse, France}
\author{Nicolas Laflorencie}
\email{nicolas.laflorencie@irsamc.ups-tlse.fr}
\affiliation{Laboratoire de Physique Th\'eorique, Universit\'e de Toulouse and CNRS, UPS (IRSAMC), F-31062, Toulouse, France}
\date{\today}

\begin{abstract}
    We present results of numerical simulations performed on one-dimensional spin chains in order to extract the so-called relaxation rate $1/T_1$ accessible through NMR experiments.
    Building on numerical tensor network methods using the \textit{Matrix Product States} (MPS) formalism, we can follow the non-trivial crossover occurring in critical chains between the high-temperature diffusive classical regime and the low-temperature response described by the Tomonaga-Luttinger liquid (TLL) theory, for which analytical expressions are known.
    In order to compare analytics and numerics, we focus on a generic spin-$1/2$ XXZ chain which is a paradigm of gapless TLL, as well as a more realistic spin-$1$ anisotropic chain, modelling the DTN material,  which can be either in a trivial gapped phase or in a TLL regime induced by an external magnetic field.
    Thus, by monitoring the finite temperature crossover, we provide quantitative limits on the range of validity of TLL theory, that will be useful when interpreting experiments on quasi one-dimensional materials.
\end{abstract}

\maketitle

\section{Introduction}\label{sec:intro}
One-dimensional ($1$d) quantum systems are known to be very peculiar due to  strong quantum fluctuations that prohibit long-range order and can give rise to unusual phases of matter. In this context, it is remarkable that quantum spin chains fall generically into two classes regarding their low-energy properties~\cite{Haldane1983,Giamarchi}: (i) critical behavior where gapless low-energy excitations can be described in the framework of Tomonaga-Luttinger liquid (TLL) theory; (ii) gapped behavior.

Nevertheless, condensed-matter experiments are mostly done on \emph{quasi}-$1$d materials, hence the role of small inter-chain couplings (as compared to the dominant 1d energy scale $J_{\rm 1d}$) may become important at low-enough temperature (eventually leading to magnetic ordering). Conversely, at high temperature ($T\gg  J_{\rm 1d}$), quantum fluctuations vanish so that a classical picture emerges. As a consequence, for realistic experimental systems the validity of a universal $1$d  TLL regime is not granted and should be checked in some unbiased way. In particular, understanding the intermediate temperature regime $T\sim J_{\rm 1d}$, highly relevant to understand several experimental data, is a great theoretical challenge regarding dynamical observables.

In this paper, we focus on nuclear magnetic resonance (NMR) for quantum spin systems~\cite{horvatic_nmr_2002}, and more specifically on the $1/T_1$ spin-lattice relaxation rate. Indeed, this quantity contains lots of information on the dynamical properties of the system since it is directly related to dynamical spin-spin correlations. Moreover, being a local quantity (a crucial property of NMR technique), we will argue that reliable data can be obtained even though we will simulate finite spin chains.

Being of fundamental interest,  the low-$T$ behavior of the NMR relaxation rate has been investigated for several 1d or quasi-1d quantum magnets. Spin-gapped compounds, such as two-leg ladders SrCu$_2$O$_3$~\cite{azuma_observation_1994}, BiCu$_2$PO$_6$~\cite{alexander_impurity_2010}, ${\mathrm{Sr}}_{14-x}{\mathrm{Ca}}_{x}{\mathrm{Cu}}_{24}{\mathrm{O}}_{41}$~\cite{magishi_spin_1998}, weakly coupled Haldane chains ${\mathrm{Y}}_{2}$${\mathrm{BaNiO}}_{5}$~\cite{shimizu_spin_1995}, or dimerized spin chains AgVOAsO$_4$~\cite{ahmed_alternating-spin-chain_2015},
exhibit an activated relaxation at low-$T$. For gapless Heisenberg chain systems, the low-energy critical behavior has been studied~\cite{Sachdev1994,Sandvik1995,Starykh1997b,Barzykin2001} for Sr$_2$CuO$_3$ which is an almost ideal realization with a large $J_{\rm 1d}\sim 2000$~K and much smaller $3$d couplings so that N\'eel temperature is pushed down to $T_N\simeq 5$~K. For such an SU(2) symmetric material, a careful comparison of experimental and numerical  NMR data has shown the prominent role of logarithmic corrections~\cite{Takigawa1997}.

Another route to TLL behavior is to apply an external magnetic field on gapped materials such as spin-$1$ Haldane gap compound~\cite{Goto2006} (CH$_3$)$_4$NNi(NO$_2$)$_3$ or dimerized spin-$1/2$ chains~\cite{Izumi2003}.
For such systems, a theoretical analysis of the $1/T_1$ behavior has been performed in Refs.~\onlinecite{Orignac2007,Klanjsek2008}.

A useful experimental review on NMR properties of several spin chains can be found in Ref.~\onlinecite{KrugvonNidda2009}. Note also that $1/T_1$ measurements have also been used to characterize one-dimensional metallic phase in carbon nanotube~\cite{Ihara2010} or quasi-$1$d superconductor~\cite{Zhi2015}.

More recently, interesting quasi-$1$d spin-gapped materials have also been investigated using NMR~\cite{Mukhopadhyay2012}:
an anisotropic spin-$1$ system NiCl$_2$-4SC(NH$_2$)$_2$ (DTN) and a spin-ladder one (C$_5$H$_{12}$N)$_2$CuBr$_4$ (BPCB). In both cases, $1/T_1$ measurements could be interpreted either as coming from magnon (respectively spinon) excitations in the gapped (respectively gapless) $1$d phase, and the quantum critical regime was also argued to be universal. Most importantly, the whole temperature range, including $1$d as well as $3$d regimes, was discussed.

Experimentally, when decreasing temperature, the NMR relaxation rate $1/T_1$ has been found to diverge in the TLL regime, with power-law governed by a characteristic exponent. Such an analysis is used in experiments to determine the corresponding TLL exponent $K$~\cite{Klanjsek2015prl,Klanjsek2015prb}. For example, it was a smoking-gun signature of \emph{attractive} TLL in (C$_7$H$_{10}$N)$_2$CuBr$_4$ (DIMPY) compound~\cite{Jeong2013,Jeong2016}. However, given that we are generically dealing with quasi-$1$d materials, critical fluctuations and $3$d ordering will limit the low-energy $1$d regime, and a genuine TLL critical behavior is observable only within some finite window in temperature. This remains to be analyzed more quantitatively, which is the main purpose of this work.

The rest of the paper is organized as follows.
In Sec.~\ref{sec:models}, we present the theoretical models and provide useful definitions. Section~\ref{sec:numerics}  describes the numerical technique based on finite temperature Matrix Product States (MPS) approach. Results are then discussed in Sec.~\ref{sec:results}. Finally, we present our conclusion in Sec.~\ref{sec:conclusion}.

\section{Models and definitions}\label{sec:models}

We give in this section the two models that will be studied in this paper and a small discussion on their phase diagram. Both models present a TLL gapless phase and a gapped phase, induced by an external magnetic field. We will also provide definition of the NMR relaxation rate $1/T_1$ and discuss its expected behavior with temperature.

\subsection{Theoretical models}

\subsubsection{The spin-$1/2$ XXZ chain}

We  first consider one of the simplest paradigmatic example of TLL liquid, namely the spin-$1/2$ XXZ chain Hamiltonian:
    \begin{multline}
        \mathcal{H_\mathrm{XXZ}} = J\sum_{j=1}^{L-1} \left(S^x_{j}S^x_{j+1}+S^y_{j}S^y_{j+1} + \Delta S^z_{j}S^z_{j+1}\right)\\
        - g\mu_Bh\sum_{j=1}^{L}S^z_{j}
        \label{xxz_ham}
    \end{multline}
where $\Delta \in (-1,1]$ denotes the Ising anisotropy, $J$ the coupling strength and $h$ is an applied magnetic field in the $z$ direction with $g$ the gyromagnetic factor and $\mu_B$ the Bohr magneton constant. The Hamiltonian is defined with open boundary conditions (OBC), as will be used in our numerical simulations.

\begin{figure}[!t]
    \centering
    \includegraphics[width=\columnwidth]{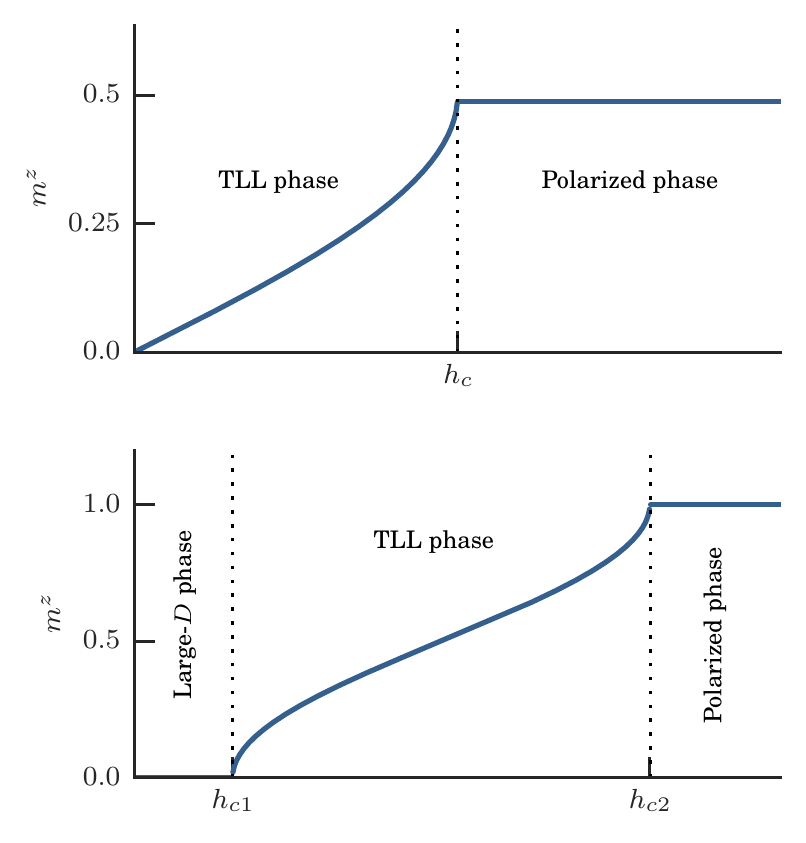}
    \caption{(color online) (i) Upper panel : magnetization curve of the XXZ Hamiltonian~\eqref{xxz_ham} as a function of the magnetic field for $\Delta \in (-1,1]$. There is a gapless TLL phase below $h_c=J(\Delta+1)/g\mu_B$ and a gapped one above when the system is fully polarized. (ii) Lower panel : magnetization curve of the $1$d DTN Hamiltonian~\eqref{dtn_ham_1d} as a function of the magnetic field. Two gapped phases (large-$D$ and polarized) are respectively located below $h_{c1}$ and above $h_{c2}$. The intermediate gapless phase can be described by a TLL theory.}
    \label{fig:magnetization_curve}
\end{figure}

In the range  $\Delta \in (-1,1]$ the XXZ model can be described by a TLL as long as its spectrum remains gapless~\cite{Giamarchi}. As a function of magnetic field, the gapless regimes extends up to a critical field $g\mu_Bh_c=J(\Delta+1)$, and the system becomes gapped for $h>h_c$. In the latter regime, the gap increases linearly with the applied magnetic field, $\Delta_g=g\mu_B(h-h_c)$, see Fig.~\ref{fig:magnetization_curve}.

\subsubsection{The quasi-$1$d spin-$1$ compound ``DTN''}\label{sec:dtn}

We also discuss a quasi-$1$d magnetic insulator compound NiCl$_2$-4SC(NH$_2$)$_2$, also called DTN, whose relevant $3$d structure consists of weakly coupled $S=1$ chains in the two other transverse (with respect to the chain axis) directions. Its experimental interest comes from the appearance of a Bose-Einstein condensation (BEC) phase when applying a magnetic field at low temperature~\cite{zapf_bose-einstein_2006,Paduan2012}. More recently, Br-doped (disordered) DTN was suggested to be a good experimental candidate for observing a Bose glass phase~\cite{Giamarchi1988,Fisher1989,Yu2012}.

Although there is 3d magnetic order observed below $T_N\sim 1$ K in DTN~\cite{zapf_bose-einstein_2006} due to weak inter-chain couplings along the two transverse directions, $J_{3\mathrm{d}}/J_{\rm 1d}\simeq 0.08$, one expects 1d physics and a TLL regime at higher $T$. The effective Hamiltonian to describe this situation reads
    \begin{multline}
        \mathcal{H_\mathrm{DTN-1d}}=J\sum_{j=1}^{L-1} \mathbf{S}_j\cdot\mathbf{S}_{j+1}+\sum_{j=1}^{L} \left[D \left(S^{z}_j\right)^2-g\mu_BhS^z_j\right],
        \label{dtn_ham_1d}
    \end{multline}
where $\mathbf{S_j}=(S^x_j, S^y_j, S^z_j)$ are spin-$1$ operators. In the current literature,~\cite{Filho2007} $J=2.2\, \mathrm{K}$ is the 1d antiferromagnetic coupling and $D=8.9\, \mathrm{K}$ is the single-ion anisotropy. The magnetic field $h$ is given in Tesla with $g=2.31$\footnote{Using this value $g=2.31$ for the g-factor for DTN, together with the most frequently used set of couplings $(D,J_{\rm 1d},J_{\rm 3d})=(8.9,2.2,0.18)$ K yields a second critical field in perfect agreement with the most precise estimates from NMR at $h_{c2}=12.32$ T~\cite{Mukhopadhyay2012,Blinder2016}.}.

The phase diagram of this 1d Hamiltonian~\eqref{dtn_ham_1d} is sketched in Fig.~\ref{fig:magnetization_curve}. In the absence of magnetic field, due to the large on-site anisotropy $D$, the system is in the so-called large-$D$ phase~\cite{Schulz1986}. This is a trivial phase, adiabatically connected to the product state $|\rightarrow\ldots\rightarrow\rangle$ where each state is in a non-magnetic $S^z=0$ eigenstate. Clearly, this phase has a finite spin-gap, which corresponds to the first critical field $h_{c1}$ needed to magnetize the system. Its value is known to be, at first order in $J/D\ll 1$~\cite{Papanicolaou1997}: $h_{c1}/g\mu_B=D-2J+\mathcal{O}(J^2/D)\simeq 3\,\mathrm{T}$. At finite magnetic field there is a gapless TLL regime for $h\in [h_{c1},h_{c2}]$, with $h_{c2}/g\mu_B=D+4J= 11.40\,\mathrm{T}$. Above this critical saturation field, the system becomes gapped again, entering a fully polarized phase. As a side remark, we recall that in the true 3d material DTN, both critical fields are shifted due to interchain couplings, so that $h_{c1}=2.10(5)\, \mathrm{T}$~\cite{wulf_critical_2015}, and $h_{c2}=12.32\, \mathrm{T}$~\cite{Blinder2016}.

In the TLL phase and close to the upper critical field $h_{c2}$, the DTN Hamiltonian~\eqref{dtn_ham_1d} can be mapped toward an effective XXZ model of spins $S=1/2$~\eqref{xxz_ham}. Using perturbation theory, effective parameters~\cite{Mukhopadhyay2012,Psaroudaki2014} are given by $\tilde{J}=2J$ and $\tilde{\Delta}=0.5$. This result can be refined using contractor renormalization (CORE) method~\cite{Morningstar1996,Capponi2004} leading to the same value of $\tilde{J}$ but a slightly reduced $\tilde{\Delta}=0.36$. Both mappings lead to a value of the effective magnetic field $\tilde{h} = h - J - D$.

\subsection{Relaxation rate $\mathbf{1/T_1}$}\label{sec:T1_def}

The nuclear spin-lattice relaxation rate $T_1^{-1}$ measured by NMR~\cite{Abragam} is basically testing the \textit{local} and \textit{dynamical} spin correlation function $S^{aa}_{nn}\left(\omega_0\right)$ with $a=x,y,z$, and $\omega_0$ being the NMR frequency at a given site $n$,
\begin{eqnarray}
    \frac{1}{T_1}&=&\frac{\gamma^2}{2}\left\{A^2_\perp \left[S^{xx}_{nn}\left(\omega_0\right)+S^{yy}_{nn}\left(\omega_0\right)\right] + A^2_\parallel S^{zz}_{nn}\left(\omega_0\right)\right\}\nonumber\\
    &=& \frac{1}{T_1^\perp} + \frac{1}{T_1^\parallel}
    \label{T1_def1}
\end{eqnarray}
Here, $A_\perp$ is the transverse hyperfine coupling constant, $A_\parallel$ the longitudinal one, $\gamma$ the gyromagnetic ratio and
\begin{multline}
   S^{ab}_{nn}\left(\omega\right)= \mathrm{Re}\left\{2\int_0^\infty\mathrm{d}t\,e^{i\omega t}\left[\langle S^a_n\left(t\right)S^b_n\left(0\right)\rangle\right.\right.\\
   -\left.\left.\langle S^a_n\left(t\right)\rangle\langle S^b_n\left(0\right)\rangle\right]\vphantom{\int_0^\infty}\right\},
   \label{dyn_corr_def}
\end{multline}
with $S^a = {S^b}^\dag$, $S^a(t)=e^{i\mathcal{H}t}S^ae^{-i\mathcal{H}t}$, and $\langle\rangle$ is the thermal average defined later in~\eqref{mps_temp_0bs}. For convenience, the $x$ and $y$ spin components can be expressed using the raising and lowering operators,
\begin{equation}
    S^{xx}_{nn}\left(\omega\right)+S^{yy}_{nn}\left(\omega\right)=\frac{1}{2}\left[S^{+-}_{nn}\left(\omega\right)+S^{-+}_{nn}\left(\omega\right)\right].
\end{equation}

It is theoretically justified to take the limit $\omega_0\rightarrow 0$ since the NMR frequency is of a few tens or hundreds of $\mathrm{MHz}$, corresponding to temperatures of the order of mK, thus being the smallest energy scale in the problem. Indeed, such temperatures are neither reached in experiments of interest nor in our numerical simulations, in particular for our purpose of probing the finite temperature TLL regime in quasi-1d systems. As a side remark, the two correlations $S^{+-}_{nn}\left(\omega_0\right)$ and $S^{-+}_{nn}\left(\omega_0\right)$ become equivalent in this limit $\omega_0 \rightarrow 0$.

The weight of the transverse and longitudinal contributions in the relaxation rate $1/T_1$ is experimentally governed by the hyperfine coupling tensors $A_\perp$ and $A_\parallel$. To favor one over the other, a specific nucleus  can be targeted for the NMR experiment. In the case of DTN, proton ${}^1$H (nuclear spin $I=1/2$) probes both components while nitrogen ${}^{14}$N ($I=1$) probes dominantly the transverse one~\cite{Blinder2016}.

Hyperfine coupling tensors put aside or set equal to one, the low temperature behavior of the transverse and longitudinal components of $1/T_1$ depends on microscopic parameters of the model. At low temperature, the transverse component is larger than the longitudinal one and it is thus justified to consider $1/T_1\simeq 1/T_1^\perp$. In the following we will fix $\gamma^2 A^2_{\perp,\parallel}=1$, and compute the relaxation rates
\begin{equation}
    \frac{1}{T_1^\perp} = \int_0^\infty\mathrm{d}t\,\mathrm{Re}\left[\langle S^\pm_n\left(t\right)S^\mp_n\left(0\right)\rangle\right],
    \label{t1_perp}
\end{equation}
and
\begin{multline}
    \frac{1}{T_1^\parallel}= \int_0^\infty\mathrm{d}t\,\mathrm{Re}\left[\langle S^z_n\left(t\right)S^z_n\left(0\right)\rangle-\langle S^z_n\left(t\right)\rangle\langle S^z_n\left(0\right)\rangle\right].
    \label{t1_parallel}
\end{multline}

Note that the single operator averages are time independent,   $\langle S^z_n\left(t\right)\rangle=\langle S^z_n\left(0\right)\rangle$.

\subsection{Tomonaga-Luttinger liquid description in the gapless phase}

The 1d gapless phase, experimentally accessible by tuning a control parameter such as the external magnetic field, can be effectively described by the TLL Hamiltonian\cite{Giamarchi},
    \begin{equation}
        \mathcal{H}_{\mathrm{TLL}}=\frac{1}{2\pi}\int \mathrm{d}r\left\{uK\left[\partial_r\theta\left(r\right)\right]^2+\frac{u}{K}\left[\partial_r\phi\left(r\right)\right]^2\right\}
        \label{ll_ham}
    \end{equation}
where $u$ is the velocity of excitations and $K$ is the dimensionless TLL parameter. They both fully characterize the low-energy properties of the system and are thus model-dependent. $\theta(r)$ and $\phi(r)$ are bosonic fields obeying the commutation relation $[\phi(x), \theta(y)]=i\pi\delta(x-y)$.

For the XXZ model \eqref{xxz_ham}, the TLL parameters $K$ and $u$ are known from Bethe ansatz equations~\cite{Luther1975}. At zero magnetic field, analytical expressions are known as a function of the Ising anisotropy $\Delta$:
    \begin{equation}
        K=\frac{\pi}{2\arccos\left(-\Delta\right)}\quad\mathrm{and}\quad \frac{u}{J}=\frac{\pi\sqrt{1-\Delta^2}}{2\arccos{\Delta}}.
        \label{ll_params_xxz}
    \end{equation}
For generic non-integrable models, the TLL parameters can be obtained numerically using DMRG by fitting static correlation functions~\cite{Hikihara2001,Hikihara2004} which has been successfully done in the past years for various quasi-$1$d compounds~\cite{Klanjsek2008,Schmidiger2012}.

In the TLL framework, the dynamical correlation $S^{ab}_{nn}(\omega)$ defined in \eqref{dyn_corr_def} can be computed analytically as a function of the temperature in the ``\textit{low energy limit}'' which we will try to define more precisely in this paper. Let us recall that the dynamical spin susceptibility is defined as,
\begin{equation}
    \chi_{ij}^{ab}\left(t\right)=-i\Theta\left(t\right)\langle \left[S^a_i\left(t\right), S^b_j\left(0\right)\right]\rangle,
    \label{susceptibility}
\end{equation}
where $\Theta(t)$ is the Heaviside function. Since a typical relevant case in experiments is the \emph{local} quantity $i=j=n$, we only consider this case in the following. In frequency space, the susceptibility can be related to the dynamical spin correlation function by~\cite{Sachdev2011}
\begin{equation}
    S^{ab}_{nn}\left(\omega\right)=\frac{2}{e^{-\beta\omega}-1}\mathrm{Im}\left[\chi_{nn}^{ab}\left(\omega\right)\right].
    \label{corr_susceptibility}
\end{equation}
In the low energy limit $\beta\omega\ll 1$ an analytical expression for Eq.~\eqref{corr_susceptibility} can be obtained, leading to~\cite{Klanjsek2008,Bouillot2011,Bouillot2011thesis}
\begin{multline}
    \frac{1}{T_1^\perp} =\\
     \frac{2A_x\cos\left(\frac{\pi}{4K}\right)}{u}\left(\frac{2\pi T}{u}\right)^{\frac{1}{2K}-1} B\left(\frac{1}{4K},1-\frac{1}{2K}\right),
   \label{ll_spsm_corr}
\end{multline}
 and
 \begin{multline}
    \frac{1}{T_1^\parallel} = \frac{A_z\cos\left(\pi K\right)}{2u}\left(\frac{2\pi T}{u}\right)^{2K-1}B\left(K,1-2K\right)\\
    +\frac{KT}{4\pi u^2},
    \label{ll_szsz_corr}
 \end{multline}
with $B(x,y)$ the Euler beta function and $A_{x,z}$ prefactors of the static correlation functions. Thus, generically $1/T_1^\perp(T)$ diverges at zero temperature as a $K$-dependent power-law, and dominates over $1/T_1^\parallel$. Note that for finite magnetic field (equivalent to non half-filled case), additional subleading corrections are expected~\cite{Suzuki2006}.

\subsection{Gapped regime behavior}
In contrast to a TLL  gapless phase where $1/T_1$ has power-law behavior at low temperature, we also consider the gapped regime where fluctuations are exponentially suppressed~\cite{jolicur_ensuremathsigma-model_1994,troyer_thermodynamics_1994,sagi_theory_1996,kishine_spin_1997,sachdev_low_1997,
Orignac2007}, such that for $T< \Delta_g$
\begin{equation}
    \frac{1}{T_1^{\perp,\parallel}} \propto \exp \left(-\alpha_{\perp,\parallel} \Delta_g/T\right),
    \label{T1_gapped_phase}
\end{equation}
where $\Delta_g$ is the energy gap of the system, and $\alpha_{\perp,\parallel}$ an $\mathcal{O}(1)$ prefactor which depends on the relaxation processes, and also on the temperature range~\cite{kishine_spin_1997}. Below, in section~\ref{sec:gappedxxz}, we show numerical results for the high-field gapped regime of the XXZ chain where our data are perfectly described by $\alpha_{\perp,\parallel}=1$. At higher temperature above the gap,  one may also expect a non-trivial crossover to TLL regime~\cite{Dora2007}.

\section{Numerical methods}\label{sec:numerics}

To get the relaxation rate one needs to obtain in the first place the dynamical correlation $\left\langle S^a_n\left(t\right)S^b_n\left(0\right)\right\rangle$. We used the TEBD (Time Evolution Block Decimation) algorithm~\cite{Vidal2004} with both real/imaginary time through the MPS formalism adapted for $1$d systems~\cite{Schollwock2011}. A general one-dimensional system containing $L$ sites with OBC can be represented by the following MPS,
\begin{equation}
    |\Psi\rangle = \sum_{\{s_i\}} A^{s_1}_{a_1}A^{s_2}_{a_1a_2}\cdots A^{s_L}_{a_{L-1}}|{s_1}\rangle|s_2\rangle\cdots |s_L\rangle
    \label{general_mps_form}
\end{equation}
where the local index $s_i$ is the physical index representing an element of the local Hilbert space at site $i$. Its dimension is $d$ and is equal to $2$ ($\uparrow$ and $\downarrow$) for spin-$1/2$ or $3$ ($\uparrow$, $\downarrow$ and $\rightarrow$) for spin-$1$. We note $a_i$ the bond index whose dimension is directly related to the ``number of states'' $m$ to describe the system, meaning that $m$ is a control parameter in the numerical simulations.

The first step is to perform an imaginary time evolution on the system to reach the desired temperature. Once the state is at hand, the second step consists of evolving it through a real time evolution. At each time step, the correlation is measured. When all the data in time-space have been obtained, a numerical Fourier Transform can be performed to get the data in frequency-space.

\subsection{Time evolution with MPS}\label{subsec:tebd_mps}

We will be general and consider the case with a Hamiltonian $\mathcal{H}$ consisting of nearest-neighbor interactions only -- it is the case for the XXZ \eqref{xxz_ham} or DTN Hamiltonian \eqref{dtn_ham_1d} introduced before. Now, we need to evolve our MPS up to a time $t$. The operation can be discretized using smaller time steps $\tau$ such that $t=N\tau$, leading to $e^{-it\mathcal{H}} = \prod^N e^{-i\tau\mathcal{H}}$. If the time step $\tau$ is small enough, a first (or higher) order Trotter decomposition can be performed,
\begin{equation}
    e^{-it\mathcal{H}} \simeq \prod^{N} e^{-i\tau\mathcal{H}_{\mathrm{even}}} \prod^{N} e^{-i\tau\mathcal{H}_{\mathrm{odd}}} + \mathcal{O}(\tau^2)
    \label{mps_trotter_tevol}
\end{equation}
where $\mathcal{H}_{\mathrm{even}}$ and $\mathcal{H}_{\mathrm{odd}}$ respectively correspond to the even and odd bond Hamiltonians only acting on two nearest-neighbor spins. The decomposition is possible because even -- odd -- bond Hamiltonians commute with each others. But it is not exact and leads to an error in $\tau$ due to the fact that $[\mathcal{H}_{\mathrm{even}}, \mathcal{H}_{\mathrm{odd}}] \neq 0$. The advantage is that the bond Hamiltonians can easily be diagonalized and exponentiated since they are only $d^2\times d^2$ matrices.

And so, applying successive evolution gates on the MPS as well as singular value decompositions to restore the MPS original tensor-site dependent form \eqref{general_mps_form} will eventually lead to a time-$t$ evolved state.

\subsection{Finite temperature with MPS}

It was useful to introduce time-evolution concepts also to discuss finite-temperature with MPS~\cite{Verstraete2004}. The main idea is to represent the density matrix $\rho_\beta$ of the physical (mixed) state in an artificially enlarged Hilbert space as a pure state $|\Psi_\beta\rangle$ -- which is what we can deal with in the MPS formalism. The auxiliary space can simply be constructed as a copy of the original one.

Assuming that we know the purification of the density matrix $\rho_{\beta=0}$ as a wave function $|\Psi_{\beta=0}\rangle$ it can be shown that an imaginary time evolution has to be performed over the infinite temperature state in order to get the finite temperature state,
\begin{equation}
    |\Psi_\beta\rangle = e^{-\beta\mathcal{H}/2} |\Psi_{\beta=0}\rangle
    \label{mps_temp_tevol}
\end{equation}
with the Hamiltonian only acting on the physical sites. This imaginary time evolution can be performed using the TEBD algorithm described before in \ref{subsec:tebd_mps}. Expectation values can then be measured at inverse temperature,
\begin{equation}
    \langle\mathcal{O}\rangle_\beta = \frac{\mathrm{Tr}\left[\mathcal{O}e^{-\beta\mathcal{H}}\right]}{\mathrm{Tr}\left[e^{-\beta\mathcal{H}}\right]}=\frac{\langle\Psi_\beta|\mathcal{O}|\Psi_\beta\rangle}{\langle\Psi_\beta|\Psi_\beta\rangle}
    \label{mps_temp_0bs}
\end{equation}
For this procedure to work, the initial state $|\Psi_{\beta=0}\rangle$ has to be a product state of Bell states between each physical site and its associated auxiliary site,
\begin{equation}
    |\Psi_{\beta=0}\rangle = \frac{1}{\sqrt{\mathcal{N}}}\prod_{n=1}^L\sum_{\{s\}}|p^n_sa^n_s\rangle
    \label{mps_temp_init}
\end{equation}
with $|p\rangle$ corresponding to physical sites, $|a\rangle$ to auxiliary ones and $\mathcal{N}$ a normalization constant. The summation is over the $d$ possible local states $s$. Such a state is simple enough to be built exactly in the MPS formalism.

\subsection{Numerical limitations}

The main limitations are about the temperature and the final time one can reach using the methods described above. The reason in both cases is directly related to a rapid growth in the entanglement entropy while evolving the state. This implies to keep larger and larger number of states $m$ in the MPS if one wants to be accurate, strongly limiting numerical simulations in practice.

On the one hand, it becomes increasingly difficult to reach low temperatures. Indeed, one expects a volume-law entanglement entropy (\textit{i.e.} linear with the system size $L$) due to the auxiliary sites which are used to purify the thermal state. As a consequence, the number of kept states $m$ needed to describe accurately the system will grow exponentially as the temperature $T$ decreases.

On the other hand, the maximal (real) time that can be reached is of the order of few tens of $J^{-1}$ typically, for similar reasons as discussed above, namely the linear growth of entanglement entropy with time~\cite{Laflorencie2016}. Thus, fixing a maximum number of kept states $m$ limits simulations to a finite time $t_{\mathrm max}$. Note that this limitation applies at \textit{all} temperatures, even $T=0$.

Despite these severe limitations, recent progress in the field has allowed some improvements. For instance, we will make use of the auxiliary degrees of freedom which are used to purify the thermal state by time-evolving them with $-\mathcal{H}$, which is mathematically exactly the same but has been shown to improve substantially the time range~\cite{Karrasch2013}. By construction, this trick only applies to finite-temperature simulations though. Last, although its use did not prove to be systematically reliable in our case, we would like to mention the possibility to use so-called \textit{linear prediction} technique, coming from data analysis~\cite{Barthel2013} which aims at predicting ``longer time'' behavior from the knowledge of dynamical correlations at ``intermediate time''.

\section{Results}\label{sec:results}

We provide in this section our numerical results~\footnote{ITensor library, \url{http://itensor.org}} about the relaxation rate $1/T_1$ using models and techniques presented in the previous sections \ref{sec:models} and \ref{sec:numerics}. First of all we will focus on the XX model (equivalent to free fermions) for which we can compute exactly the dynamical correlations for all temperatures and that will serve as a benchmark for our simulations. Next, we will turn to the interacting XXZ case for $S=1/2$, and then to a $S=1$ chain model relevant to the DTN material.

\subsection{Case study : XX point ($\Delta=0$)}
\begin{figure*}[!ht]
    \centering
    \includegraphics[width=2\columnwidth]{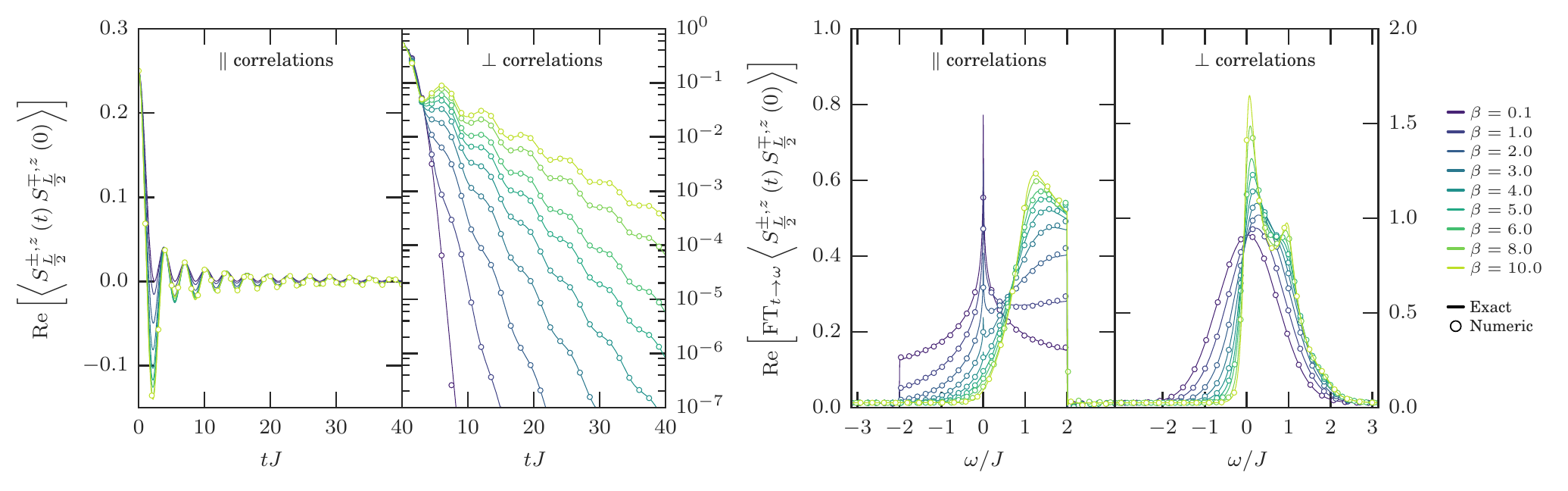}
    \caption{(color online) We compare numerical results (circles) used to determine the $1/T_1$ with analytical results (straight lines) for the XX model. For the transverse ($\perp$) case, exact results are computed on a chain of size $L=64$ (OBC). As for the longitudinal ($\parallel$) case, the chain size is $L=1000$ (OBC). Numerics on their side are performed on a chain of $L=64$ (OBC) sites. The left panel shows the real value of the dynamical correlations. For readability, we only display numerical results for the lowest temperature for the longitudinal correlations. Indeed, this is a priori the hardest to compute and thus the most subject to errors. The right panel shows the real part of the Fourier transform of the real time data. Although we only show data up to $t=40~J^{-1}$ the Fourier Transform of the exact $zz$ correlations was performed using data up to $t=1000~J^{-1}$.}
    \label{fig:xx_twspace}
\end{figure*}

The XXZ Hamiltonian \eqref{xxz_ham} at $\Delta=0$, known as XX model, can be mapped onto a model of free spinless fermions using a Jordan-Wigner transformation. We restrict ourselves to $h=0$. It can be diagonalized in Fourier space with $\varepsilon_k=J\cos k$ and $k=\frac{n\pi}{L+1}$ with $n=1, 2,\ldots,L$ considering open boundary conditions,
    \begin{equation}
        \mathcal{H}_{\mathrm{XX}}=\frac{J}{2}\sum_{j=1}^{L-1}\left(c^\dag_jc_{j+1} + \mathrm{h.c.}\right) = \sum_k \varepsilon_k c^\dag_k c_k.
    \end{equation}
Unlike the bosonization expressions \eqref{ll_spsm_corr} and \eqref{ll_szsz_corr} which are only valid in the low-energy limit, the results presented in this section will be valid for \emph{all} regimes. We present the details of the calculations in appendix~\ref{app:xx} for the analytical exact expressions of the dynamical correlations for the XX model using precisely the same conditions as in our numerical simulations (\textit{i.e.} a finite chain length with open boundary conditions).

We show the `bare' results in Fig.~\ref{fig:xx_twspace} that will be used to obtain the relaxation rate thereafter. Ideally, one is interested in the thermodynamic limit (\textit{i.e.} $L\rightarrow\infty$) but we see that, at finite temperature (hence finite correlation length), working on finite length chains with only a moderate number of sites $L$ allows to get reliable data. Indeed, the MPS estimates agree perfectly with the exact expressions (see appendix~\ref{app:xx}).

First of all, we consider the local dynamical correlation of the site in the middle of the chain reducing de facto boundary effects. Then, as finite size effects are known to be caused by the reflection of the propagating excitations at TLL velocity $u$ on the boundary of the system, one can estimate a time below which the dynamical correlations can be considered as free of finite size effects (basically, $ut\sim L$).

We  first discuss the transverse correlations, see Fig.~\ref{fig:xx_twspace}. For all temperatures, they decay rather quickly to zero, so that we can safely truncate data to a maximum time $t_{\mathrm{max}}$ (which is anyway a natural cutoff provided by the inverse of the NMR frequency $\omega_0$) and get reliable values of $1/T_1^\perp$ by integrating over time. Moreover, we have also checked that finite size effects are extremely small since we are computing a local correlation.

The same cannot be said for the longitudinal correlations. They continue to oscillate even for high temperatures and long times, and their amplitude gets (very) slowly smaller with time. This implies severe limitations to get data in the thermodynamic limit. For instance, exact computations using \eqref{long_corr_xx} were done on $L=1000$ and still displayed oscillations of amplitude around $10^{-4}$ at $t=1000~J^{-1}$. This makes the value of $1/T_1^\parallel$ very difficult to estimate. This well-known behavior is related to spin diffusion-like behavior~\cite{Fabricius1998,Sirker2006} which cause a logarithmic divergence at small frequency $\omega$. However, we have to remember that the NMR frequency $\omega_0$ eventually provides a natural cutoff.

For completeness we display the real part of the Fourier transform on the right panels of Fig.~\ref{fig:xx_twspace} for which the $1/T_1$ value as defined in section~\ref{sec:T1_def} corresponds to the $\omega=0$ value.

\subsection{Spin-1/2 XXZ chain at $\Delta\neq 0$}
\subsubsection{Gapless regime}
\begin{figure}[t]
    \centering
    \includegraphics[width=\columnwidth]{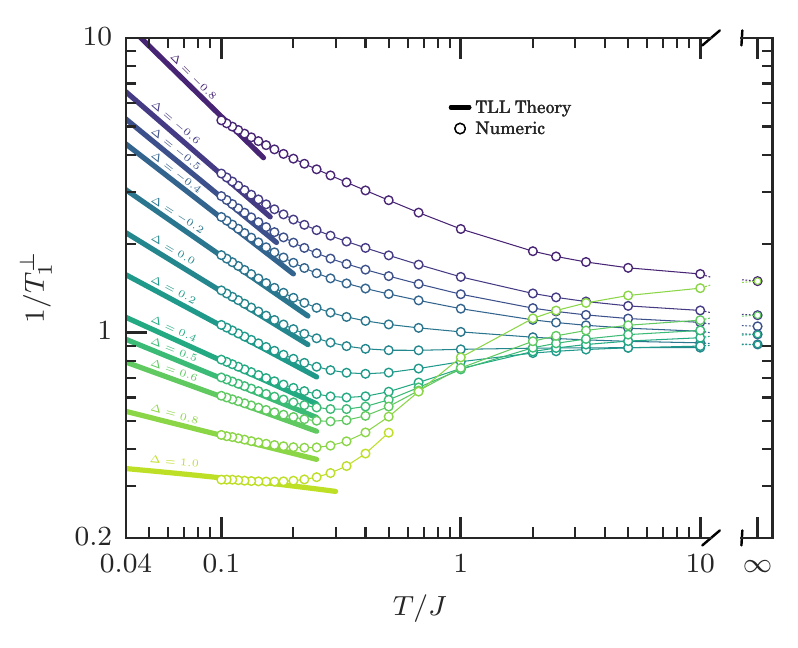}
    \caption{(color online) Transverse relaxation rate $1/T_1^\perp$ {\it{vs.}} reduced temperature $T/J$ for the spin-$1/2$ XXZ chain at various $\Delta$ and $h=0$ obtained numerically using MPS techniques (circles, from top to bottom: $\Delta=-0.8,\,-0.6,\,-0.5,\,-0.4,\,-0.2,\,0.0,\,0.2,\,0.4,\,0.5,\,0.6,\,0.8,\,1$). Numerics are compared to TLL theory Eq.~\eqref{ll_spsm_corr} at low temperature (thick lines) for $|\Delta|<1$, and with Eq.~\eqref{eq:su2} for the SU(2) Heisenberg point $\Delta=1$. The thin lines between the circles are guides to the eyes.}
        \label{fig:xxz_t1_perp}
\end{figure}

Building on the perfect agreement observed previously between MPS estimates and the exact analytical solution of the XX model, we are now confident to extend our study of the more generic XXZ case $-1<\Delta\leq 1$, described by a TLL, and compute the relaxation rates. Results are plotted in Fig.~\ref{fig:xxz_t1_perp} for various values of the anisotropy. The simulations were performed on systems of size $L=64$ with a cutoff of $\varepsilon=10^{-10}$ in the singular values. We kept a maximal number of $D=500$ states. A fourth order Trotter decomposition was used with a Trotter step of $\tau=0.1$.

First, in the gapless regime we do observe an excellent quantitative agreement between numerical estimates and the TLL prediction Eq.~\eqref{ll_spsm_corr} at low-enough temperature.  This asymptotic regime with a power-law behavior $\sim T^{\frac{1}{2K}-1}$ occurs only below $T/J\sim 0.1-0.2$ (depending on the anisotropy $\Delta$). Here we stress that \emph{there are no free parameters in the analytic expressions}. Indeed, the TLL parameters are computed using the exact expressions~Eq.~\eqref{ll_params_xxz} for $u$ and $K$, and $A_x$ is obtained following Refs.~\onlinecite{Lukyanov1997,Lukyanov1999}. The isotropic limit $\Delta=1$ is a special point where logarithmic corrections appear in several quantities~\cite{affleck_critical_1989,eggert_susceptibility_1994,affleck_exact_1998}, leading to a very slow divergence of the (isotropic) NMR relaxation rate~\cite{barzykin_temperature-dependent_2000,Barzykin2001}
\begin{equation}
\frac{1}{T_1}\simeq\frac{1}{\sqrt{2\pi^3}}\sqrt{\ln\frac{\Lambda}{T}+\frac{1}{2}\ln\left(\ln\frac{\Lambda}{T}\right)},
\label{eq:su2}
\end{equation}
where $\Lambda\simeq 24.27J$. MPS estimates compare well with this \emph{parameter-free} expression, as visible in Fig.~\ref{fig:xxz_t1_perp}.

Interestingly, we notice the non-monotonic behavior of $1/T_1^\perp$ with temperature only when $\Delta\gtrsim 0$ (which corresponds to repulsive or vanishing interactions in the fermionic language).

As a last comment, we have observed that for infinite temperature ($\beta=0$), the value of $1/T_1^\perp$ does not depend on the sign of $\Delta$, which is expected since the many-body spectrum of ${\mathcal{H}}_\Delta$ is an odd function of $\Delta$. Its value is minimum for $\Delta=0$ with $1/T_1^\perp=\sqrt{\pi}/(2J)$~\cite{Fabricius1997} and increases with $|\Delta|$. At the isotropic point $|\Delta|=1$ we expect the relaxation rate to diverge due to the diffusion-like behavior~\cite{Fabricius1998,Sirker2006} of the dynamical correlation function. Our results at infinite-$T$ go beyond Baker-Campbell-Hausdorff expansion developed up to ${\cal O}(t^2)$ in Ref.~\onlinecite{Moriya1956} to compute $\langle S^{\pm}_j(t)S^{\mp}_j(0)\rangle$ at short times, which would suggest $1/T_1^\perp\sim J^{-1}(1+\Delta^2)^{-\frac{1}{2}}$. This prediction is in contrast to what we found, namely the transverse relaxation rate \emph{increasing} with $|\Delta|$. Indeed, while such an expansion finds the correct  gaussian behavior for $\Delta=0$ (free-fermions), higher-order terms have to be taken into account for $|\Delta|>0$ where the transverse dynamical correlation function at longer times gets larger when increasing $|\Delta|$.

\subsubsection{Gapped XXZ chain}
\label{sec:gappedxxz}
\begin{figure}[t]
    \centering
    \includegraphics[width=\columnwidth]{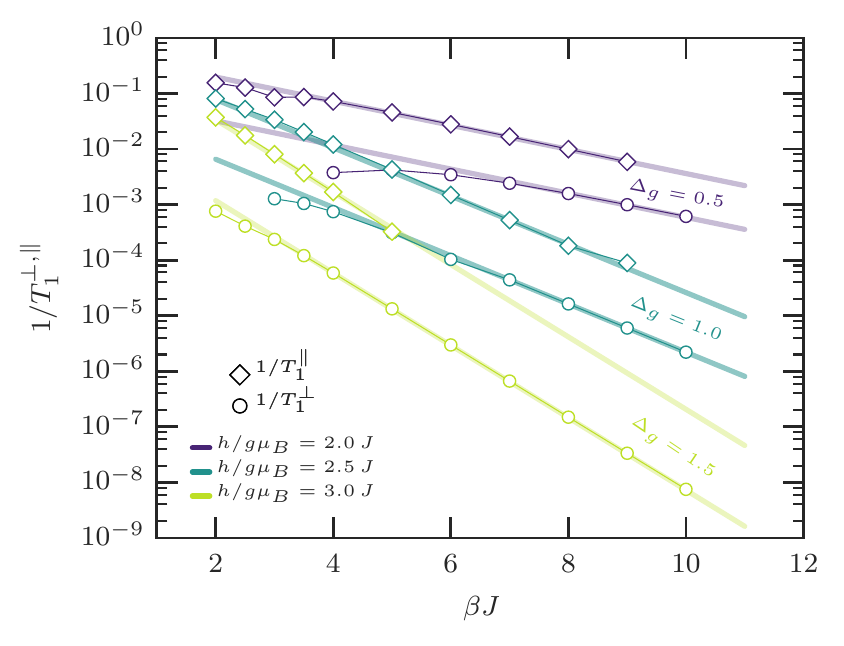}
    \caption{(color online) Transverse and longitudinal relaxation rates $1/T_1^{\perp,\parallel}$ plotted against reduced inverse temperature $\beta J$ for the spin-$1/2$ XXZ chain in its gapped phase for the anisotropy $\Delta=0.5$. The critical magnetic field is $h_c=3J/2g\mu_B$ and the value of the gap $\Delta_g=g\mu_B\left(h-h_c\right)$. Numerical results are obtained using MPS techniques (circles and diamonds) and the exponentially decaying behavior is verified with the straight lines set with the expected gap value $1/T_1^{\perp,\parallel}=c_{\perp,\parallel}\cdot e^{-\beta\Delta_g}$ and $c_{\perp,\parallel}$ a non-universal free parameter.}
    \label{fig:xxz_t1_perp_delta_0.5}
\end{figure}

We then set the anisotropy value to $\Delta=0.5$ and apply a magnetic field to move into the gapped phase. Transverse and longitudinal relaxation rates $1/T_1^{\perp,\parallel}$ are plotted in Fig.~\ref{fig:xxz_t1_perp_delta_0.5} where we observe an excellent agreement with an exponentially activated behavior $\sim \exp(-\beta\Delta_g)$, where $\Delta_g $ is the spin gap. We notice that as the gap gets smaller, the lower the temperature has to be to observe the exponential law.

\subsection{DTN}

We now move to the DTN compound in its $1$d limit described by Eq.~\eqref{dtn_ham_1d}. We compute the relaxation rates for various values of the magnetic field $h$, mainly close to $h_{c2}$ which is relevant for NMR experiments~\cite{Mukhopadhyay2012}. It is a more challenging system to simulate than the XXZ model as it is made of spins $S=1$ (enlarged local Hilbert space). The simulations were performed on open chains of size $L=64$ with a cutoff of $\varepsilon=10^{-10}$ in the singular values. We kept a maximal number of $D=150$ states. A fourth order Trotter decomposition was used with a Trotter step of $\tau=0.02$.

\begin{figure}[t]
    \centering
    \includegraphics[width=\columnwidth]{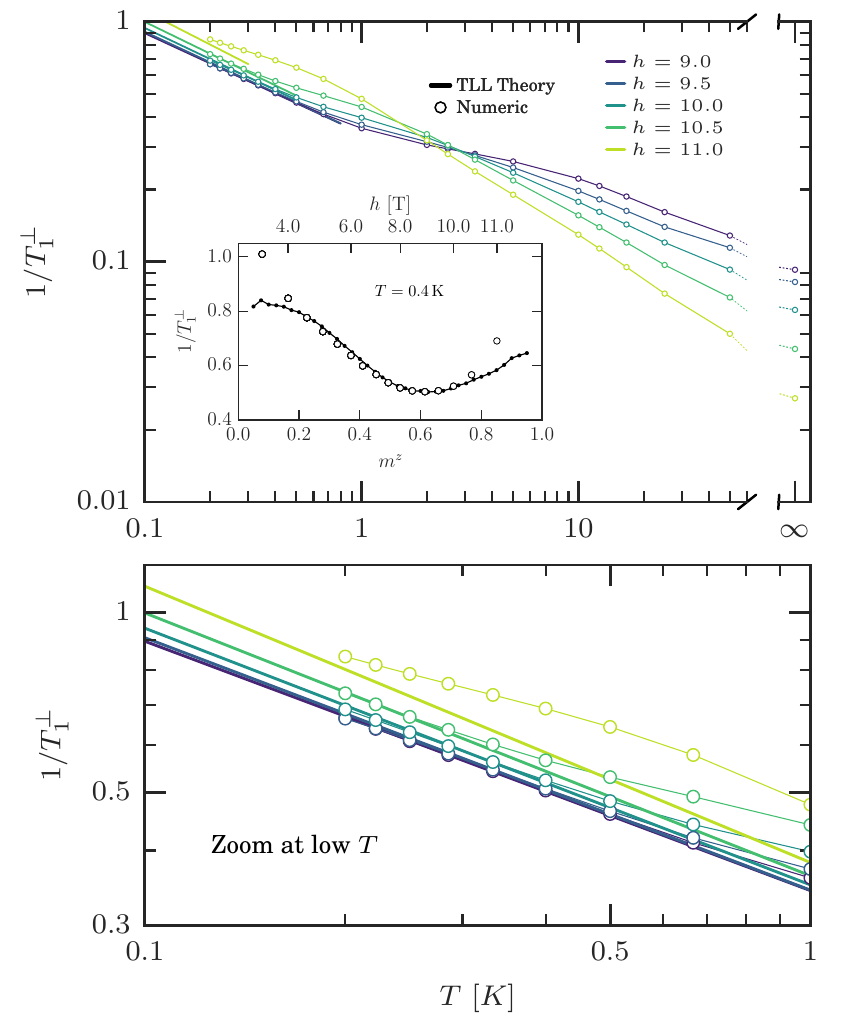}
    \caption{(color online) Transverse relaxation rate $1/T_1^\perp$ plotted {\it{vs.}} temperature $T$ for the spin-$1$ DTN chain obtained numerically using MPS techniques (circles). The low temperature behavior is compared to TLL prediction (straight lines). The magnetic field $h$ is given in Tesla. The inset compares TLL prediction and numerical results for $T=0.4\,\mathrm{K}$ and covers the whole TLL phase from $h_{c1}$ to $h_{c2}$. The lower panel is a zoom on the low temperature asymptotic power-law regime.}
    \label{fig:dtn_t1_perp}
\end{figure}

Numerical results, shown in Fig.~\ref{fig:dtn_t1_perp}, compare extremely well with TLL prediction at low temperature. Interestingly, the TLL power-law behavior starts at slightly higher temperature, as compared to the XXZ model, $T\simeq 0.5\,\mathrm{K}$ ($T/J\sim 0.2$), especially as we approach the middle of the TLL phase, away from the critical field $h_{c2}$. We point out that there are again no adjustable coefficients, the TLL parameters being computed independently using standard DMRG~\footnote{See Ref.~\onlinecite{Blinder2016} for TLL parameters dependence on the magnetic field.}. The tiny difference that appears at low temperature between numerical data and TLL is due to the limited number of states $m$ kept when performing calculations. Though this does not dispute the TLL prediction, it reveals the challenge in such time-dependent simulations. The inset in Fig.~\ref{fig:dtn_t1_perp} shows the transverse relaxation rate at $T=0.4\,\mathrm{K}$ for various values of the magnetic field covering the whole range from $h_{c1}$ to $h_{c2}$. Once more, there is a very good agreement between numerics and TLL theory except when one gets close to the critical fields. Indeed, as we clearly see in the lower panel of Fig.~\ref{fig:dtn_t1_perp} for $h=11.0\,\mathrm{T}$, the power law is not met yet for the lowest temperature we could reach $T=0.2\,\mathrm{K}$.

The non-monotonic behavior of $1/T_1^\perp$ observed in the XXZ model is absent for the DTN and may seem odd at first place since it can be mapped effectively onto a $S=1/2$ XXZ chain with $\Delta=0.5$ or $0.36$ and could thus be compared with Fig.~\ref{fig:xxz_t1_perp}. However this non-monotonic variation is observed at high temperature while this mapping is only justified in the low-energy limit as discussed in~\ref{sec:dtn}.

One can also try to compare the relaxation rates of Fig.~\ref{fig:dtn_t1_perp} with the NMR data for the DTN compound given in Ref~\onlinecite{Mukhopadhyay2012}. What draws our attention is the non-monotonic regime of $1/T_1^\perp$ observed at high temperature experimentally, which, as we have just discussed, is not theoretically predicted for a single DTN chain. Yet it cannot be attributed to $3$d effects as $J_{3\mathrm{d}}=0.18$ K is very small compared to the temperature $T$. We then observed that experiments are performed by proton ($^1H$) NMR which probes {\it{both}} $1/T_1^{\perp}$ and $1/T_1^{\parallel}$.

\begin{figure}[t]
    \centering
    \includegraphics[width=\columnwidth]{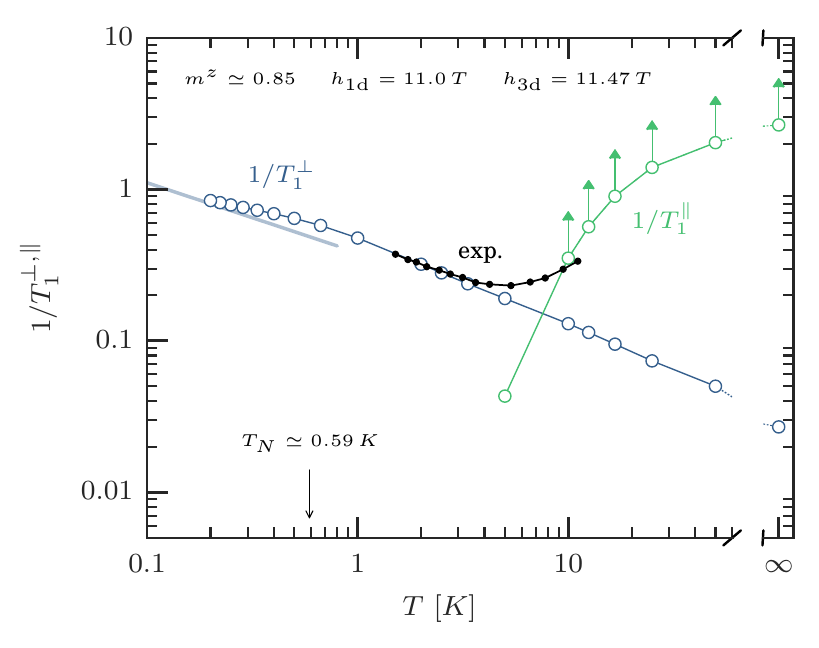}
    \caption{(color online) Longitudinal $1/T_1^\parallel$ and transverse $1/T_1^\perp$ relaxation rates for the DTN spin-$1$ chain at $h=11.0\,\mathrm{T}$, corresponding to $m^z\simeq 0.85$. As $1/T_1^\parallel$ cannot be estimated for sure, we only provide a lower bound. The non-monotonic behavior observed experimentally at high $T$ in Ref~\onlinecite{Mukhopadhyay2012} apparently comes from the large contribution of $1/T_1^\parallel$ at high temperature. Experimental data for DTN~\cite{Mukhopadhyay2012} at the same magnetization are shown for comparison, after a proper rescaling in order to match the low-$T$ regime. The 3d BEC transition temperature $T_N(m^z\simeq 0.85)\simeq 0.59$ K~\cite{Blinder2016} is also shown.}
    \label{fig:dtn_t1_para}
\end{figure}

We therefore interpret this effect as due to the parallel contribution of the relaxation rate. We show in Fig.~\ref{fig:dtn_t1_para} both the transverse and longitudinal $1/T_1^{\perp,\parallel}$ as a function of temperature. We cannot precisely estimate the value of $1/T_1^\parallel$ due to its dependence on $\omega_0$ (and therefore on our maximum time in numerical
simulations) so that we give a lower bound. Its high temperature contribution to the total relaxation rate clearly dominates over the transverse part and explains well the experimental non-monotonic regime at high $T$.

Perhaps more importantly, as displayed in Fig.~\ref{fig:dtn_t1_para}, the 3d BEC ordering observed in DTN~\cite{zapf_bose-einstein_2006,Blinder2016} for $m^z\simeq 0.85$ at $T_N\simeq~0.59\,\mathrm{K}$ occurs {\it{above}} the asymptotic regime where the genuine TLL power-law behavior is expected. It is therefore impossible to directly extract TLL exponents in DTN, because of interchain effects that eventually lead to an ordering of the coupled TLLs.  Ideally we would expect for quasi-1d systems the TLL description of the NMR relaxation to be valid in the following temperature regime: $J_{\rm 1d}\gg T\gg J_{\rm 3d}$.

Concerning the difficulty to obtain reliable data at  high temperature for the longitudinal  $1/T_1^\parallel$, it is well known that this is due to spin diffusion-like behavior~\cite{Fabricius1998,Sirker2006}. Therefore, measurements should in principle depend explicitly on the NMR frequency $\omega_0$.

\section{Conclusion}\label{sec:conclusion}

Performing time-dependent numerical simulations at finite temperature on $1$d systems to compute the NMR relaxation rate $1/T_1$, we have discussed the temperature range validity of analytical predictions for two models (i) the paradigmatic example for Tomonaga-Lutinger liquids: the spin-$1/2$ XXZ chain for various Ising anisotropies, and (ii) a more realistic $S=1$ Hamiltonian, relevant for experiments on the DTN compound as a function of an external magnetic field.

Both models present in some regime a gapless phase that can be described by TLL ``\textit{low-energy}'' theory,
 with a relaxation rate dominated by its transverse component $1/T_1^\perp\sim T^{\frac{1}{2K}-1}$ algebraically diverging at low temperature, where $K$ is the dimensionless TLL exponent. We observed that the expected power-law behavior occurs only below $T/J_{1\mathrm{d}}\sim 0.1-0.2$, thus defining the low-energy limit of validity of TLL theory an order of magnitude below the energy scale $J_{1\mathrm{d}}$ of the system. It is important to be able to define this limit as TLL predictions are often used experimentally on quasi-1d compounds to extract the value of $K$.
As a consequence, we believe that it remains experimentally challenging~\cite{Izumi2003}, and often impossible, to explore a genuine critical $1$d regime  in quasi-$1$d compounds when $J_{1\mathrm{d}}$ is small and $3$d ordering prevents a wide TLL regime. For instance, we have shown that for DTN, the BEC ordering temperature is larger than the crossover temperature towards TLL behavior.

We have also studied the transverse relaxation rates of these two models in other regimes than TLL theory. First, we considered high temperatures, with a peculiar non-monotonic behavior in the $S=1/2$ XXZ model in the repulsive regime at high $T$, which does not exist for the 1d $S=1$ model of DTN. However, such a non-monotonic dependence with temperature at high $T$ is experimentally observed in DTN. We showed that this effect comes from the parallel contribution of the relaxation rate $1/T_1^\parallel$ dominating at high temperature over the transverse part. Finally, we verified that in gapped phases the relaxation rates are exponentially suppressed $\sim\exp\left(-\Delta_g/T\right)$ and can indeed lead to accurate determinations of the spin gap, at least in a regime of temperature $T\ll\Delta_g$ since other relaxation mechanisms can change the activated behavior at higher temperature.

We want to emphasize again the role of $3$d ordering at finite temperature, preventing  the observation of a $1$d TLL regime. As discussed for the particular case of DTN, one needs a hierarchy of energy scale $J_{\rm 1d}\gg T\gg J_{\rm 3d}$ to be able to directly extract the TLL exponent $K$ from the divergence of $T_1^{-1}$ with $T$.

Concerning future advances for quasi-1d systems, we can envision trying to simulate imaginary-time dynamics using quantum Monte-Carlo techniques, provided that the model has no minus-sign problem. While it will be necessary to perform a numerical analytic continuation (using for instance Maximum Entropy techniques), we have some hope that this could lead to reliable results to quantitatively capture the influence of interchain effects. For 1d chains, it has been rather successful~\cite{Sandvik1995}. As a matter of fact, our extensive $1$d results could serve as useful benchmarks for that too.

While we have considered various chains, we are far from being exhaustive. Indeed, there are some TLL models for which elementary excitations may not be simple spin flips, for instance multipolar nematic phase for which $1/T_1$ behavior will be different~\cite{Sato2009,Sato2011}.
In a similar line of thought, we could imagine simulating more complicated models including charge and spin degrees of freedom to describe NMR relaxation in metallic or superconducting wires. \\

{\it Note added~:} While completing this work, a related numerical study by Coira {\it et al.} has appeared~\cite{Coira2016}. Our results are perfectly compatible with each other when comparison can be made, such as the transverse $1/T_1$ data for a single spin-$1/2$ XXZ chain with $\Delta\geq 0$.\\

\section*{Acknowledgment}

The authors are grateful to M. Horvati\'c and M. Klanj\v{s}ek for their careful reading of this manuscript and thoughtful comments and to R. Blinder, T. Giamarchi and E. Orignac for valuable discussions. This work was performed using HPC resources from GENCI (Grant No. x2015050225 and No. x2016050225), and is supported by the French ANR program BOLODISS and R\'egion Midi-Pyr\'en\'ees.
\appendix
\section{Dynamical correlations for the XX model}
\label{app:xx}
For completeness, we remind the reader the exact expressions for time-displaced spin correlations in the exactly solvable XX model.
\subsection{Longitudinal correlations}

The longitudinal correlations $\langle S^{z}_i\left(t\right)S^{z}_j\left(0\right)\rangle$ for the XX model are basically density correlations when performing the Jordan-Wigner transformation. Since we are interested in the temperature dependence of the correlations, what we need to compute is actually,
\begin{multline}
    \langle S^{z}_i\left(t\right)S^{z}_j\left(0\right)\rangle =\frac{1}{\mathcal{Z}}\mathrm{Tr}\left[e^{i\mathcal{H}t}S^{z}_ie^{-i\mathcal{H}t}S^{z}_je^{-\beta\mathcal{H}}\right]
\end{multline}
with $\mathcal{Z}=\mathrm{Tr}e^{-\beta\mathcal{H}}$ the partition function. Calculations lead to\cite{Katsura1970},
\begin{equation}
    \langle S^{z}_i\left(t\right)S^{z}_j\left(0\right)\rangle = \frac{1}{4}I_1 I_2
    \label{long_corr_xx}
\end{equation}
with $I_1$,
\begin{multline}
    I_1 = \frac{2}{L+1}\sum_{k}
    \sin ki\sin kj\left[1 + \tanh\frac{\beta\varepsilon_k}{2}\right]e^{-i\varepsilon_k t}
\end{multline}
and $I_2$,
\begin{multline}
    I_2 = \frac{2}{L+1}\sum_{k}
    \sin ki\sin kj\left[1 - \tanh\frac{\beta\varepsilon_k}{2}\right]e^{+i\varepsilon_k t}.
\end{multline}
Also, the single operator averages are
\begin{equation}
    \langle S^z_i(t)\rangle = \langle S^z_i(0)\rangle = 1 - \frac{4}{L+1} \sum_k \frac{\sin^2 ki}{e^{-\beta\varepsilon_k}+1}.
\end{equation}

\subsection{Transverse correlations}

The transverse correlations $\langle S^{\pm}_i\left(t\right)S^{\mp}_j\left(0\right)\rangle$ have a more complicated structure in the fermion representation due to the string of operators in the exponential\cite{Stolze1995}. We introduce the following identity $e^{i\pi c^\dag_lc_l} = (-1)^{c^\dag_lc_l} = A_lB_l$ with $A_l=c^\dag_l+c_l$ and $B_l=c^\dag_l-c_l$, leading to
\begin{multline}
    2\langle S^{\pm}_i\left(t\right)S^{\mp}_j\left(0\right)\rangle = \left\langle\left[\prod_{l=1}^{i-1}A_l(t)B_l(t)\right]A_i(t)\right.\\
    \left.\left[\prod_{l=1}^{j-1}A_l(0)B_l(0)\right]A_j(0)\right\rangle.
\end{multline}
Now thanks to Wick's theorem this product of many fermion operators can be rewritten as elementary expectation values of two operators through the Pfaffian of some skew-symmetric matrix. Its elements above the diagonal (which is purely made of zeros) are
\begin{equation}
    \begin{matrix}
        \langle A_1(t)B_1(t)\rangle& \langle A_1(t)A_2(t)\rangle&\dots&\langle A_1(t)A_j(0)\rangle\\
        &\langle B_1(t)A_2(t)\rangle&\dots&\langle B_1(t)A_j(0)\rangle\\
        &&\dots&\dots\\
        &&&\langle B_{j-1}(0)A_j(0)\rangle\\
    \end{matrix}
\end{equation}
At finite temperature,
\begin{equation}
    \langle A_i(t)B_j(0)\rangle =  \frac{1}{\mathcal{Z}}\mathrm{Tr}\left[e^{i\mathcal{H}t}A_ie^{-i\mathcal{H}t}B_je^{-\beta\mathcal{H}}\right].
\end{equation}
The two-body expectation values can be computed going in Fourier space,
\begin{multline}
    \langle A_i(t)A_j(0)\rangle=\frac{2}{L+1}\sum_k \sin kj \sin ki\\
    \left[\cos\varepsilon_k t-i\sin\varepsilon_k t\tanh\frac{\beta\varepsilon_k}{2}\right]
\end{multline}
and,
\begin{multline}
    \langle A_i(t)B_j(0)\rangle=\frac{2}{L+1}\sum_k \sin kj \sin ki\\
    \left[-\sin\varepsilon_k t-i\cos\varepsilon_k t\tanh\frac{\beta\varepsilon_k}{2}\right].
\end{multline}
And equivalently $\langle B_i(t)B_j(0)\rangle=-\langle A_i(t)A_j(0)\rangle$ as well as $\langle B_i(t)A_j(0)\rangle=-\langle A_i(t)B_j(0)\rangle$.

\end{document}